\documentclass[conference]{IEEEtran}

\usepackage{amsmath,amsfonts,amssymb,amsthm}
\usepackage{epsfig,url}
\usepackage{graphicx,color,footmisc}
\usepackage{subcaption}
\usepackage{cite}
\usepackage{tfrupee} 
\usepackage{tikz}
\usepackage{adjustbox}
\usepackage{bytefield}
\usepackage{dblfloatfix} 
\usepackage{multicol}
\usepackage{multirow}

\title{Realizing the Frugal 5G Network}
\author{\IEEEauthorblockN{Meghna Khaturia\IEEEauthorrefmark{1}, Pranav Jha\IEEEauthorrefmark{1} and Abhay Karandikar\IEEEauthorrefmark{1}\IEEEauthorrefmark{2} }
	\IEEEauthorblockA{\IEEEauthorrefmark{1}Department of Electrical Engineering, Indian Institute of Technology Bombay, Mumbai, India}
	\IEEEauthorblockA{\IEEEauthorrefmark{2}Director and Professor, Indian Institute of Technology, Kanpur, India}}
   
\begin{document}
\maketitle
\begin{abstract}
    In order to make effective use of the Internet, broadband connectivity is a pre-requisite. However, in the majority of rural areas in developing countries, high-speed connectivity is unavailable. The Frugal 5G network architecture presented in this paper aims at enabling broadband in rural areas by addressing the challenges associated with it. The work presented in this paper is a development over our previous work in \cite{frugal5g} \& \cite{khaturiaefficient}, in which we proposed abstract network architecture for Frugal 5G. In this paper, we provide an innovative solution to realize the Frugal 5G network. We identify the key system requirements and show that the proposed solution enables an uncomplicated and flexible realization of the Frugal 5G network. We are currently building a testbed to implement the proposed changes.
\end{abstract}
\section{Introduction}

        The Internet is one of the most significant inventions of $20$th century, providing us with numerous benefits. However, around $52$\% of the global population is unconnected and is unable to benefit from its advantages~\cite{ICT2017}. The majority of the unconnected population lives in rural areas of developing countries. These areas have limited coverage of optical fiber/digital subscriber line; the technologies typically used to provide broadband connectivity. In such a scenario, cellular broadband technology, such as 4G, seems to have the potential to enable the Internet connectivity in such places. However, due to challenges, such as low average revenue per user, thin population, and irregular supply of electricity, cellular deployment in these areas is also limited. Even the upcoming 5th Generation (5G) mobile technology standards, with a focus on 1 Gbps data rate, 500 km/h mobility and 1 ms latency, may not be geared towards addressing the challenges of rural broadband connectivity~\cite{5G}. Thus, there is an urgent need to develop a suitable solution, which can enable Internet connectivity in rural/remote areas of developing countries.
        
        In the existing literature~\cite{karlsson2016energy,khalil2017feasibility,chiaraviglio20165g}, authors have tried to address different aspects of this problem, however no comprehensive solution has been proposed. In our previous work~\cite{frugal5g} \& \cite{khaturiaefficient}, we followed a bottom-up approach to  analyze and identify the challenges of rural connectivity and developed an \textit{abstract} network architecture to address the problem systematically and holistically. The proposed network architecture is referred to as the Frugal 5G network. In this paper, we build on our previous work and propose a method to realize the Frugal 5G network. The main contributions of the work presented in this paper are as follows.
       $\bullet$ We investigate the key system requirements for the Frugal 5G network realization.
            $\bullet$ We develop a novel mechanism to realize the proposed Frugal 5G network architecture. We show that the proposed mechanism facilitates a simple and flexible implementation of the Frugal 5G network.
            $\bullet$  We also identify additional use cases, which can be supported by this novel solution.

        
        The rest of the paper is organized as follows. In Section II, we briefly describe the rural broadband connectivity challenges and the Frugal 5G Network Architecture, as proposed in our earlier work. For further details concerning the network architecture, the reader is referred to~\cite{frugal5g}. Section III identifies the system requirements for the Frugal 5G network. In Section IV, we propose the solution to address the system requirements and effectively realize the Frugal 5G network. We conclude by discussing about our ongoing work on Frugal 5G testbed implementation.

    \section{Frugal 5G Network Architecture - Overview}
         \begin{figure*}[!ht]
                    \begin{subfigure}[b]{\textwidth}
                        \centering
                	    \includegraphics[width=9cm,trim={2.2cm 3cm 2cm 2cm},clip]{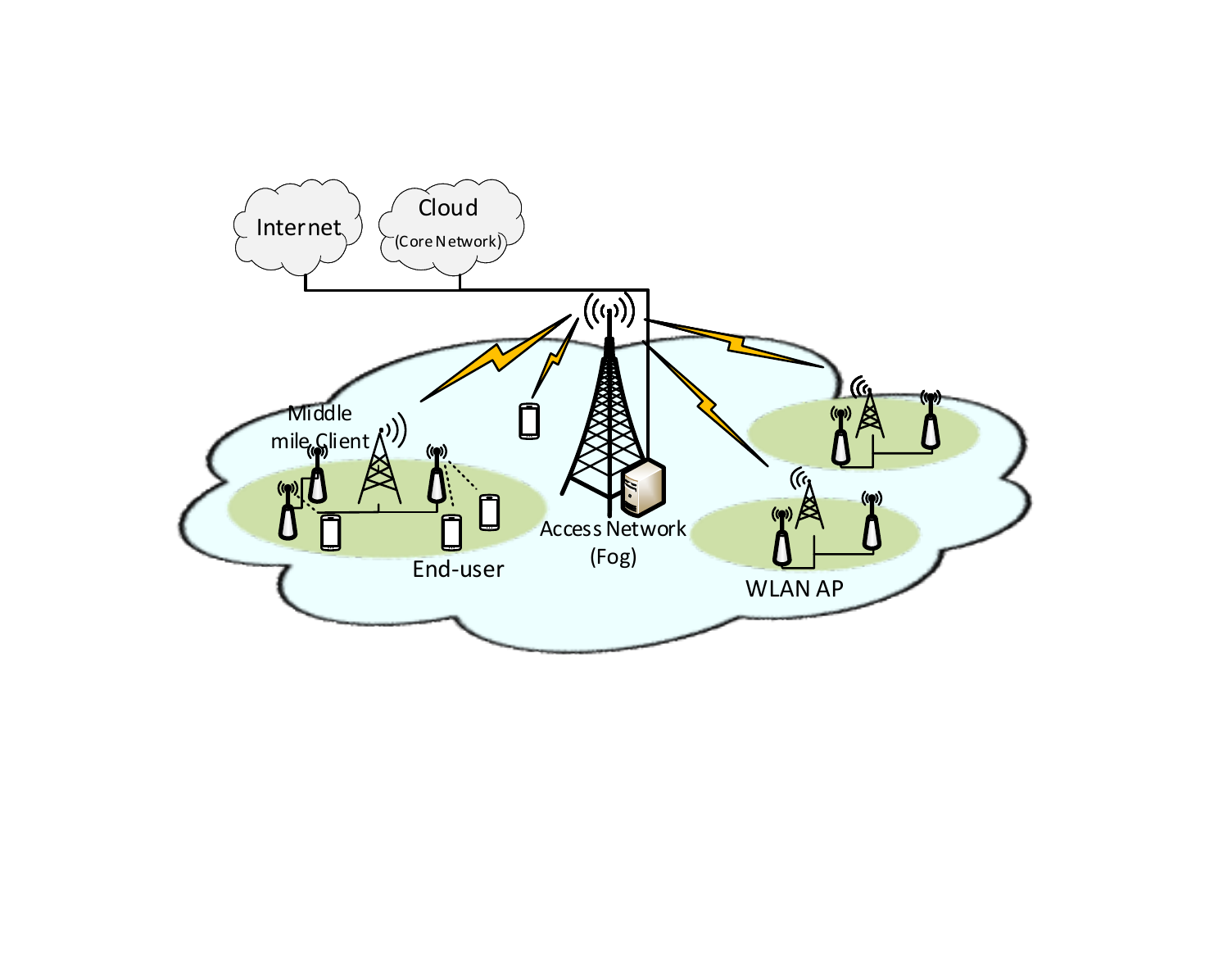}
                	    \caption{A deployment example of Frugal 5G architecture.}
                	    \label{fig:dep}
                    \end{subfigure}
                    \quad \quad
                   \begin{subfigure}[b]{\textwidth}
            		    \centering
            		    \includegraphics[height = 7cm, trim={3.5cm 2.3cm 0.5cm 0.5cm},clip]{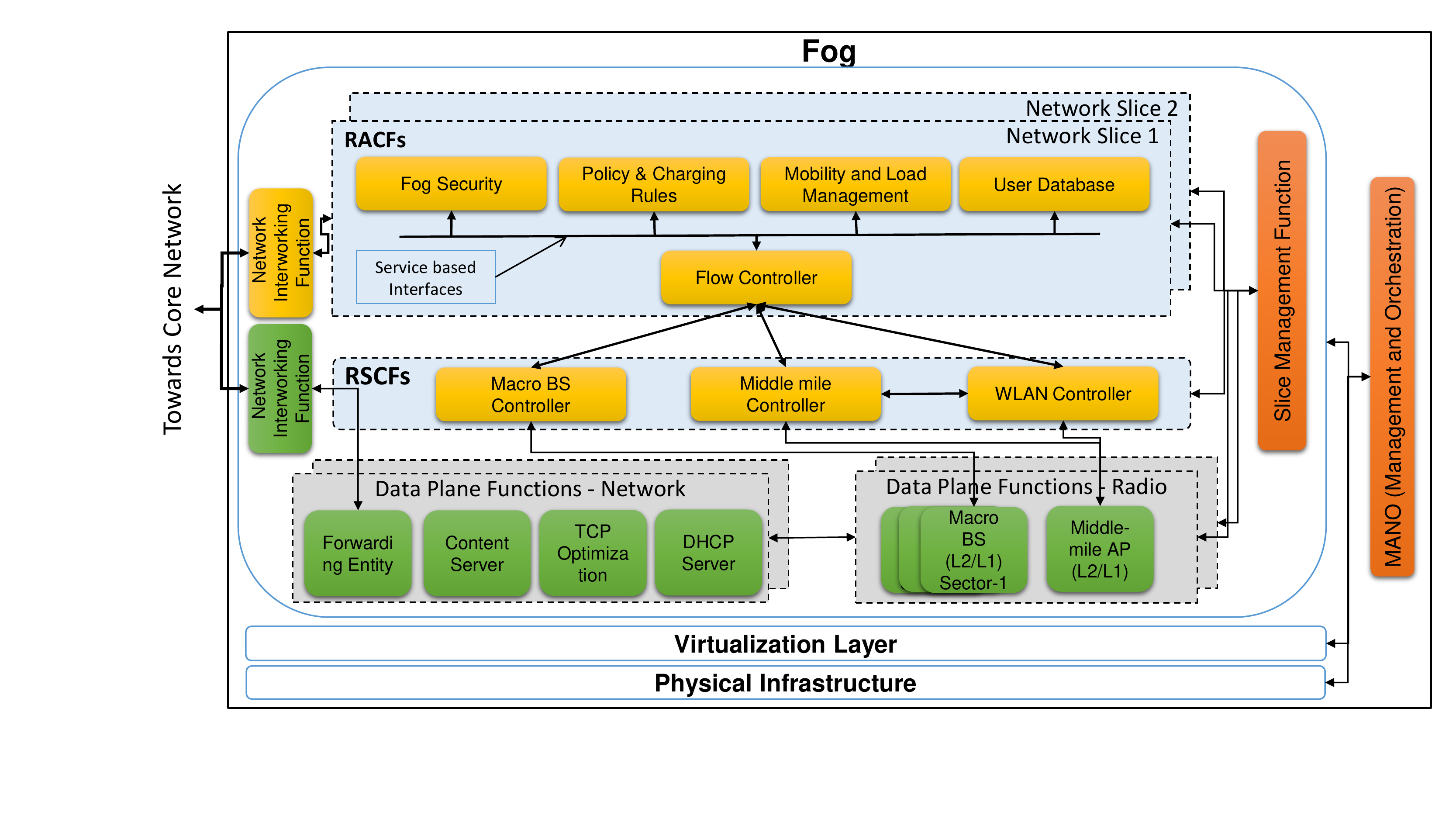}
            		    \caption{\label{fig:arch} Detailed architecture of Access Network (fog)}
                    \end{subfigure}
                    \caption{Frugal 5G Network}
       \end{figure*}
   In our previous work \cite{frugal5g}, we have identified various characteristics and challenges related to rural connectivity such as i) low population density \& clustered settlement, ii) strong community bond among villagers, iii) unavailability of relevant content and need for facilitation of region specific content generation and storage, iv) erratic supply of electricity from grid, v) low income per household and vi) no/little importance of high-speed mobility support.

   To address the challenges stated above, we have proposed an innovative design for the Frugal 5G network architecture. The Frugal 5G Access Network (AN) provides last-mile access over wireless medium to people living in rural areas as shown in Fig.~\ref{fig:dep}. To enable the Internet connectivity, the Frugal 5G AN can be connected to a standard cellular Core Network (CN) or a fixed broadband network, similar to a public Wi-Fi network~\cite{publicwifi}.  The Frugal 5G AN consists of two RATs, i.e., Wireless Local Area Network (WLAN) and macro Base Station (BS). WLANs provide high-speed connectivity inside the rural settlements, whereas macro BS provides carpet coverage in a large area, comprising of village clusters and open spaces in between. The Frugal 5G network is connected to the cellular CN/fixed broadband network through a fiber Point of Presence (PoP). Macro BS is directly connected to the PoP. However, in order to carry the traffic from the PoP to the WLAN Access Points (APs), we employ a wireless middle mile network. Optimally designing the wireless middle mile network is also a challenging problem which we have addressed in our previous work~\cite{khaturiaefficient}.
    
    To control and manage the multiple Radio Access Technologies (RATs) in the network and to meet the objectives of Frugal 5G network, we have proposed an Software Defined Networking (SDN) and Network Function Virtualization (NFV) based control framework for Frugal 5G AN as shown in Fig~\ref{fig:arch}. The controller is located at the edge of the network i.e. in the fog. It is a layered SDN controller. The top-level functions, referred to as the RAT Agnostic Control Functions (RACFs), have a unified view of the Radio Access Network (RAN) and based on this, the network decisions are taken. The RAT Specific Control Functions (RSCFs) provide an abstract view of the underlying network to the RACFs. Since the Frugal 5G network needs to support local communication within AN, AN can operate in a stand-alone manner. The SDN Controller is able to setup the data paths for local communication. When connectivity to the mobile core is available, CN performs all the standard tasks. The AN and CN are synchronized periodically to avoid any inconsistencies.  The Frugal 5G Network also has an interworking function, which acts as the interface to the CN.
    
   In summary, Frugal 5G network architecture addresses the rural connectivity challenges through its innovative features---unified control framework with SDN, usage of wireless middle-mile, the capability of isolated operation and localized communication, flexibility in service/function instantiation in the fog and the framework for network virtualization and network slice creation.
   
\section{System Requirements \& Constraints}
    In order to realize the Frugal 5G network architecture, we need to understand the system requirements and constraints. These are discussed next.
    
    \subsubsection{Cost-effective High-speed Connectivity}
    WLANs need to enable cost-effective high-speed connectivity in village clusters and a User Equipment (UE) shall be able to access the Internet over the high-speed links offered by it. 

    \subsubsection{Support for Large Cells}
    As part of Frugal 5G network, there is also the need to provide basic data connectivity in a large coverage area. It supplements the high speed links provided by WLANs, and is especially required in open areas around the villages, where WLANs may not be available. Such connectivity can be provided through deployment of macro BSs, which can be facilitated through one of the many possible technologies, e.g., 3rd Generation Partnership Project's Long Term Evolution (3GPP LTE), 5G New Radio (NR), WiMax, and IEEE 802.22. It is important that the proposed solution has the flexibility to enable usage of any one of these technologies and also supports easy migration from one technology to another. 

    \subsubsection{Flexibility in the Internet Access}
    Today, mobile UEs usually access the Internet via two mechanisms, either through a cellular network (e.g., with the help of the LTE technology) or via a fixed broadband network (typically used along with public Wi-Fi access). To have flexible connectivity, the Frugal 5G AN must be able to connect to both the cellular CN and the fixed broadband network. This would enable a mobile UE connected to Frugal 5G network to access the Internet through any one of these options. Obviously, this would also depend on the capability and the access rights of a UE.
    
    \subsubsection{Unified Multi-RAT Control}
    As mentioned above, Frugal 5G AN is a multi-RAT network and it may be possible to serve the UEs over two separate radio links, i.e., via the macro cell and/or WLAN. Therefore, it is important to enable a unified multi-RAT control and management structure in the AN so as to optimally route the data flows based on the service requirements of UEs, the network availability, and the load on the different network nodes. A unified control may help in additional network optimization tasks, e.g., energy saving, by driving a subset of WLAN and middle mile nodes in sleep mode during low traffic scenario.
    
    \subsubsection{Unified Interface towards Core/Fixed Broadband Network}
    Even though Frugal 5G network is a multi-RAT network, it supports a unified interface towards the cellular core as well as the fixed broadband network. A unified interface towards cellular core/fixed broadband network simplifies the inter-working with these networks. A unified interface also facilitates a unified control of the multi-RAT AN. 
   
    \subsubsection{Cross-network Mobility}
    UE mobility from Frugal 5G network to a standard 4G/5G network is an important requirement to enable seamless network coverage. The challenge is to introduce minimum or no changes in the UE protocol stack to support such mobility. 
    
   \begin{figure*}[!t]
	\centering
	\includegraphics[width=0.8\linewidth]{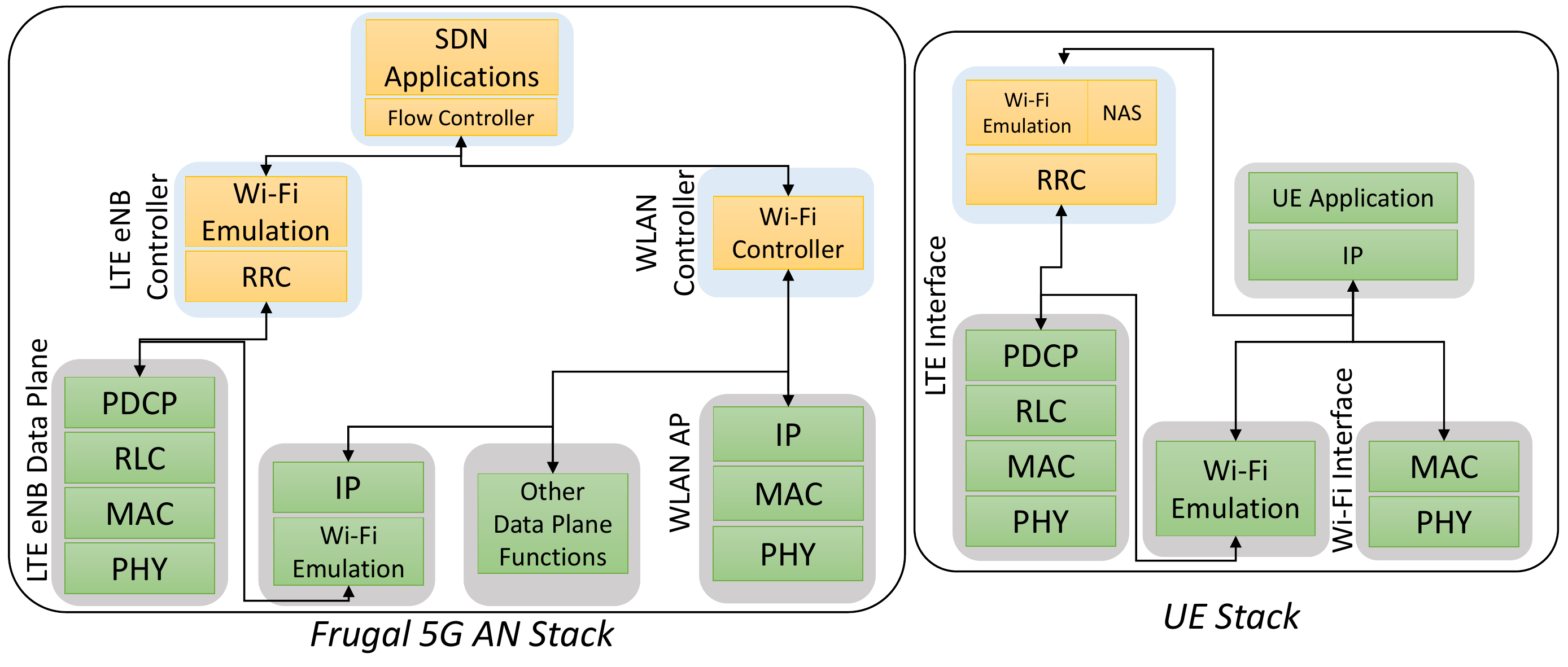}
	\caption{Proposed Modifications in the Radio Protocol Stack}
	\label{fig:stack}
\end{figure*}
\section{Proposed Realization of Frugal 5G Network}

Considering the above system requirements and constraints, we propose the following solution to realize the Frugal 5G network. As part of the solution, we identify a few candidate technologies, which have been discussed next. We also propose some changes in network as well as UE stack. 

    \subsection{Suitable Technologies and Associated Challenges}
    \subsubsection*{RAT Selection for Macro BS}
    To realize the Frugal 5G network, we propose to use the LTE standard based RAT for macro BS. LTE is a mature cellular standard with a broad deployment base and hence a cost-effective choice for the macro cell deployment in rural areas. Due to a broad deployment base, a large percentage of UEs already support this technology. When used in a sub-GHz band, it fulfills the fundamental objective of providing carpet coverage over a large area in an energy-efficient and cost-effective manner~\cite{lteenergy}. It also provides support for Quality of Service. In future, macro BSs based on other technologies such as 5G NR can also be used. 
    
    \subsubsection*{RAT Selection for WLAN}
    A suitable option for WLAN is IEEE 802.11~\cite{802.11} based Wi-Fi as it is the most commonly used WLAN technology and cost-effective and energy efficient too~\cite{wifienergy}. Most of the UEs today support IEEE 802.11 based Wi-Fi.
    
    \subsubsection*{Core Network Alternatives}
    We propose to connect 3GPP 5G CN with the Frugal 5G AN. 3GPP 5G is the most advanced cellular communication standard and it treats different RATs, both the 3GPP and non-3GPP ones, e.g., LTE and Wi-Fi, in an identical and unified manner. 

    \subsubsection*{Challenges} Certain challenges surface once we select the RATs for the Frugal 5G network. We have proposed to use LTE based eNodeB (eNB) for macro BS. However, a standard LTE eNB requires strict co-ordination with LTE CN for handling the UE requests. Therefore, connectivity to a standard LTE CN becomes mandatory to provide services under Frugal 5G network if an LTE eNB is used as the macro BS. It also means that a Frugal 5G UE connecting through the LTE eNB would not be able to access the Internet if LTE CN is not available. However, as has been mentioned above, the Frugal 5G AN may connect to a fixed broadband network only or in certain scenarios may not be connected to any external network and only provide localized communication. Additionally, one of the essential working principles of the Frugal 5G network is a unified control of the multi-RAT AN. The SDN controller in fog is responsible for setting up the flows and route data through the Frugal 5G AN based on the load and the traffic condition in AN. If a standard LTE network is used, then this would not be possible as the data traffic would flow independently over the WLAN and LTE. Even if we migrate from LTE eNB to 5G NR based gNB, the problem remains the same.
    \subsection{Principle Aspects of the Solution}
    
    In order to address the system requirements discussed above, we propose the following:\\
    $\bullet$ A standard LTE eNB comprises of three interfaces, i.e., radio interface towards the UE, S1 Interface towards the LTE CN and X2 Interface towards other eNBs. We propose to remove S1 and X2 interfaces from the eNB and use the eNB radio interface in the Wi-Fi emulation mode.\\
     $\bullet$  Wi-Fi emulation along with the removal of the S1/X2 interface stacks, makes an LTE eNB appear as a Wi-Fi AP to other entities in the network. The UE also perceives LTE radio interface as an independent Wi-Fi interface with the help of the Wi-Fi emulation.\\
     $\bullet$  With these changes, the Frugal 5G AN appears as a Wi-Fi network (a non-3GPP access) to the 5G CN/fixed broadband network. The usage of LTE eNB in AN is completely transparent to the 5G CN. The UEs under the coverage of Frugal 5G network appear as Wi-Fi users as far as the 5G CN/fixed broadband network is concerned. \\
    $\bullet$  A UE is authenticated in a unified manner irrespective of the RAT it is using to access the Internet. It may be authenticated using 802.1x based authentication procedure in the Frugal 5G network. If a UE connects to 5G CN, it can be authenticated using the standard 5G Non-Access Stratum (NAS) procedures.

     \begin{figure*}[!ht]
    \centering
    \includegraphics[scale=0.4]{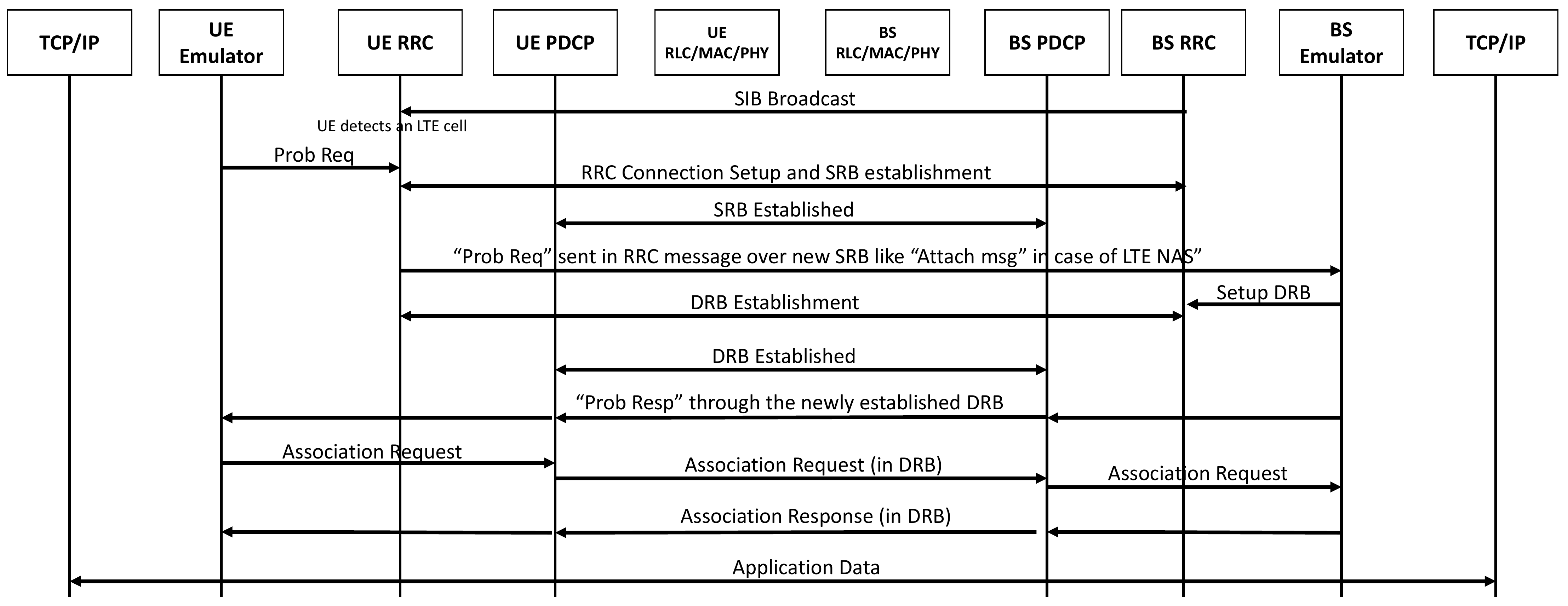}
    \caption{An Example Call Flow}
    \label{fig:callflow}
\end{figure*}
    
    \subsection{Network Side Modifications}
    
    To enable the working of the LTE radio interface in Wi-Fi emulation mode, we add a layer over the Radio Resource Control (RRC) and the Packet Data Convergence Protocol (PDCP) layer of the LTE eNB as shown in Fig.~\ref{fig:stack}. This layer is referred to as the Wi-Fi emulation layer. It sends/receives Wi-Fi Medium Access Control (MAC) frames over the LTE radio interface, i.e., over the LTE RRC or PDCP, Radio link control (RLC), MAC, and Physical (PHY) layers. It also provides an interface to the higher layers, e.g., the IP layer to send/receive Wi-Fi MAC Protocol Data Units (PDUs) containing the UE data. It extracts the MAC PDUs from the received MAC frames over the LTE radio interface and delivers it to the higher layers. Similarly, it receives data packets from the higher layers (MAC Service Data Units (SDUs)), creates Wi-Fi MAC PDUs and sends the PDUs through the LTE radio interface to the UE. The Wi-Fi Emulation layer also uses the eNB RRC layer to establish data bearers between the UE and the eNB over the LTE interface. The detail of how Wi-Fi data frames are carried over the LTE radio interface is detailed out in Section~\ref{sec:call}, where we discuss a call flow example.
    
    Owing to the Wi-Fi emulation layer, the RAT agnostic flow controller gets a uniform view of the Frugal 5G network as a Wi-Fi network, which makes the decision-making process at the flow controller uncomplicated. As both the RATs behave as Wi-Fi interfaces, the Frugal 5G network can be treated as a non-3GPP AN by the 5G CN. It also enables a unified interface towards the CN. In order to deliver the required Quality of Service (QoS) to the UEs, the Wi-Fi emulation layer also invokes the appropriate radio bearer at the LTE interface.

    \subsection{Modifications in the UE Stack}
    It is important to note that the NAS layer of the standard UE LTE stack tries to establish communication with the LTE CN. Since the LTE CN is not present, NAS layer is not used while the UE is connected to the Frugal 5G Network. Instead, the Wi-Fi Emulation layer takes its place, which is added over the RRC layer and the LTE data plane stack. Wi-Fi emulation layer uses the RRC layer to establish data bearers between a UE and an eNB through the LTE protocol stack. The emulation layer acts similar to the NAS layer as far as the RRC stack is concerned. Once the data bearers are established between a UE and an eNB, the emulation layer uses them to send the Wi-Fi MAC PDUs, carrying the UE data received from the higher layers to the peer entity (eNB). In order to enable a UE to access 5G CN, a few enhancements as defined by 3GPP~\cite{23.501} would also need to be added (not covered in this paper).

    
    
    \subsection{Sending Wi-Fi Broadcast Information} 
    It was explained in the previous sections that the unicast data (as part of the Wi-Fi MAC PDUs) are sent over the LTE interface through dedicated bearers created for the individual UEs. However, there are some broadcast Wi-Fi PDUs as well, such as beacon frames, which need to be sent to all UEs. To make LTE eNB behave as Wi-Fi AP, sending beacon frames is necessary. Beacon frames inform the UE about the presence of the Frugal 5G network which enables the Wi-Fi emulation layer. If UE doesn't detect the beacon frame, NAS layer is enabled and UE behaves like a standard UE. This feature introduces an easy method to enable cross-network mobility. We can also send push notifications to a UE when it is in sleep mode through beacon frames. 
    
    To send beacon frames, we use Multimedia Broadcast Multicast Services (MBMS) Point to Multipoint Radio Bearer (MRB). An eNB periodically transmits MBMS control information via SIB13 and Multicast Control Channel (MCCH) which helps in setting up of MRB. Once the MRB is set up, we can periodically send the beacon frames through it. Using MRB (instead of LTE broadcast channel) to send beacon frames is beneficial as no changes are made in UE protocol stack. 

\subsection{Call Flow to Send Wi-Fi PDUs over LTE Interface}
\label{sec:call}
As shown in Fig.~\ref{fig:callflow}, Wi-Fi Emulation layer generates a Probe Request which is sent over SRB to the eNB instead of Attach message. This request is received by the RRC layer in the eNB stack and sent to the Wi-Fi emulation layer. Since we have sent Probe Request (instead of Attach) over SRB, identification, authentication, security and location update procedures do not take place. In response to Probe Request, Wi-Fi Emulation layer requests the RRC layer to send RRC Connection Reconfiguration message to the UE which has the Probe Response and also activates a Data Radio Bearer (DRB). The UE then responds to eNB with RRC Connection Reconfiguration complete message. Now, the Wi-Fi emulator at UE sends the Association Request in DRB and receives the Association Response. Now, data can be transferred over LTE interface.

\section{Conclusion}
In this paper, we have proposed an innovative solution to realize the Frugal 5G network. Based on the proposed solution, a Frugal 5G test bed is being implemented in our lab. We have implemented the key functions of 5G CN, the Interworking Functions (between the CN and RAN), the RAT Agnostic SDN controller (Flow controller), the Wi-Fi controller and the Wi-Fi Emulation Layer. The network functions have been implemented in an Open Stack based dockerized environment. An LTE eNB running on TI platform and a Raspberry Pi based UE is being modified as per the proposed architecture. A Raspberry Pi based Wi-Fi AP is also a part of the testbed. We aim to present the experimental results in our future work once the testbed is ready. The solution enables an easy path of migration from LTE to 5G NR and would facilitate a unified multi-RAT RAN comprising of 5G NR and Wi-Fi technologies. It is also possible to connect to 4G Core instead of the 5G Core with ease by just changing the interworking function in the Fog. As specified in \cite{23.402}, it would be similar to how non-3GPP access connects to 4G Core. The proposed Multi-RAT RAN with a unified control structure may prove advantageous over the 3GPP 5G architecture, where different access technologies, e.g., 5G NR and Wi-Fi access are controlled independently.

The proposed Frugal 5G network solution can also be adopted to cost-effectively implement a private network, e.g. in a university campus. Seamless connectivity throughout the campus can be enabled using the proposed architecture. Frugal 5G network architecture may prove to be useful in situations of disaster as it can operate in a standalone manner.



\begin{thebibliography}{9}

\bibitem{frugal5g}
M.~Khaturia, P.~Jha, and A.~Karandikar, ``Connecting the unconnected: Towards
frugal 5g network architecture and standardization,'' \emph{arXiv preprint
	arXiv:1902.01367}, 2019.

\bibitem{khaturiaefficient}
M.~Khaturia, K.~Appaiah, and A.~Karandikar, ``{On Efficient Wireless Backhaul
	Planning for the “Frugal 5G” Network},'' in \emph{IEEE WCNC Future
	Networking Workshop for 5G and Beyond Testbed and Trials}, April 2019.

\bibitem{ICT2017}

\emph{{ICT Facts and Figures 2017}}, [Date accessed 25.05.2019]. [Online].
Available:
\url{https://www.itu.int/en/ITU-D/Statistics/Documents/facts/ICTFactsFigures2017.pdf}


\bibitem{5G}
\emph{{3GPP TR 21.915, Release 15 Description}}, 2019, (Release 15).

\bibitem{karlsson2016energy}
A.~Karlsson, O.~Al-Saadeh, A.~Gusarov, R.~V.~R. Challa, S.~Tombaz, and K.~W.
Sung, ``{Energy-efficient 5G deployment in rural areas},'' in \emph{2016 IEEE
	12th International Conference on Wireless and Mobile Computing, Networking
	and Communications (WiMob)}.\hskip 1em plus 0.5em minus 0.4em\relax IEEE,
2016, pp. 1--7.

\bibitem{khalil2017feasibility}
M.~{Khalil}, J.~{Qadir}, O.~{Onireti}, M.~A. {Imran}, and S.~{Younis},
``{Feasibility, architecture and cost considerations of using TVWS for rural
	Internet access in 5G},'' in \emph{2017 20th Conference on Innovations in
	Clouds, Internet and Networks (ICIN)}, March 2017, pp. 23--30.

\bibitem{chiaraviglio20165g}
L.~{Chiaraviglio}, N.~{Blefari-Melazzi}, W.~{Liu}, J.~A. {Gutierrez}, J.~{Van
	De Beek}, R.~{Birke}, L.~{Chen}, F.~{Idzikowski}, D.~{Kilper}, J.~P. {Monti},
and J.~{Wu}, ``{5G in rural and low-income areas: Are we ready?}'' in
\emph{2016 ITU Kaleidoscope: ICTs for a Sustainable World (ITU WT)}, Nov
2016, pp. 1--8.

\bibitem{publicwifi}

B.~Wang, K.~Wan, D.~Allan, and D.~Thorne, \emph{{TR-321--public Wi-Fi access in
		multi-service broadband networks}}, [Date accessed 21.09.2019]. [Online].
Available:
\url{https://www.broadband-forum.org/technical/download/TR-321.pdf}


\bibitem{lteenergy}
R.~{Martinez Alonso}, D.~{Plets}, M.~{Deruyck}, L.~{Martens}, G.~{Guillen
	Nieto}, and W.~{Joseph}, ``{TV White Space and LTE Network Optimization
	Toward Energy Efficiency in Suburban and Rural Scenarios},'' \emph{IEEE
	Transactions on Broadcasting}, vol.~64, no.~1, pp. 164--171, March 2018.

\bibitem{802.11}
\emph{{IEEE Std 802.11-2016 (Revision of IEEE Std 802.11-2012)}, title={IEEE
		Standard for Information technology—Telecommunications and information
		exchange between systems Local and metropolitan area networks—Specific
		requirements - Part 11: Wireless LAN Medium Access Control (MAC) and Physical
		Layer (PHY) Specifications}}, pp. 1--3534, Dec 2016.

\bibitem{wifienergy}
P.~{Serrano}, A.~{Garcia-Saavedra}, M.~{Hollick}, and A.~{Banchs}, ``{}on the
energy efficiency of ieee 802.11 wlans,'' in \emph{2010 European Wireless
	Conference (EW)}.

\bibitem{23.501}
\emph{{3GPP TS 23.501, System Architecture for the 5G System}}, 2018, (Release
15).

\bibitem{23.402}
\emph{{3GPP TS 23.402,Universal Mobile Telecommunications System (UMTS); LTE;
		Architecture enhancements for non-3GPP accesses}}, 2018, (Release 15).

\end{thebibliography}
\end{document}